# Bifocal Fresnel lens based on the polarization-sensitive metasurface

Hen Markovich, Dmitrii Filonov, Ivan Shishkin, and Pavel Ginzburg

*Abstract*—Thin structured surfaces allow flexible control over propagation of electromagnetic waves. Focusing and polarization state analysis are among functions, required for effective manipulation of radiation. Here a polarization sensitive Fresnel zone plate lens is proposed and experimentally demonstrated for GHz spectral range. Two spatially separated focal spots for orthogonal polarizations are obtained by designing metasurface pattern, made of overlapping tightly packed cross and rod shaped antennas with a strong polarization selectivity. Optimized subwavelength pattern allows multiplexing two different lenses with low polarization crosstalk on the same substrate and provides a control over focal spots of the lens only by changing of the polarization state of the incident wave. More than a wavelength separation between the focal spots was demonstrated for a broad spectral range, covering half a decade in frequency. The proposed concept could be straightforwardly extended for THz and visible spectra, where polarization-sensitive elements utilize localized plasmon resonance phenomenon.

H. Markovich, I. Shishkin, D. Filonov and P. Ginzburg are with School of Electrical Engineering, Tel Aviv University, Tel Aviv, 69978, Israel (www.tau.ac.il)

(e-mail: henmarkovich@mail.tau.ac.il (corr. author), ivanshishkin@post.tau.ac.il, dmitriif@mail.tau.ac.il, pginzburg@post.tau.ac.il ).

D. Filonov and P. Ginzburg also are with ITMO University, St. Petersburg, 197101, Russia (www.ifmo.ru)

H. Markovich, I. Shishkin, D. Filonov contributed equally.

## I. INTRODUCTION

Control over propagation of electromagnetic waves with thin subwavelength patterned surfaces is required in a broad range of applications. This approach enables significant reduction of both volumes and weights of devices used in wireless communication links. Impedance surfaces and related approaches are usually utilized for those purposes [1].

A recent interest in structured surfaces with carefully designed patterns was inspired by a series of works on optical metasurfaces (e.g. [2]–[4],[5]), which allow flexible control over light propagation with nanometer-thin layers. Subwavelength elements, forming a metasurface pattern, utilize localized plasmon resonance phenomenon, which allows achieving desired properties by means of geometrical tuning of nanoantennas' shapes (e.g. [6], [7]). Essential requirement to be met for achieving a broad range of functions is the ability to obtain a full phase control upon an interaction with individual elements (or small arrays of those).

Recent progress in optical imaging with metasurfaces along with development of new concepts in field manipulation with thin elements inspired development of new systems, operating at microwave regimes. Radio frequency (RF) imaging techniques, widely employed in traditional radar applications [8], are utilized nowadays for advancing volumetric scanning in a cluttered environment [9]. For example, 3D images of a crowd in public places are acquired at 30-300GHz frequency ranges for security reasons, e.g. [10]. Flat elements, capable of fast data analysis are required for performing the data processing. Furthermore, specular and diffusive reflections from irregular surfaces may result in a loss of information. Consequently, a significant coverage of stereo angles with detecting systems is required. While bulky and mechanically movable antennas cannot be used for these purposes from practical standpoints, cheap, flat, and integrable elements should be used. Lenses are essential elements, required for standard imaging techniques (as a remark, computational imaging could relax this demand, e.g. [11]).

Microwave lensing is intensively studied over the last 50 years [12]. Flat lenses are usually made of either single [13] or multi layered sheets with metal patterns with [14], [15] or without [16], [17] electric connection between them. Negative index metamaterials are also can be employed [18]. Recently, metasurface-based lenses on flexible substrates have been demonstrated [19]. A significant benefit of metasurface-based approach (for all frequency ranges) is in the minimized thickness of resulting structures, which found its use in variety of applications, like formation of focal spots at different angles for different polarizations [20], a single focal spot for different wavelengths [21] or correction of chromatic aberrations [22]. Therefore, the axial stacking of multiple layers is needed to achieve the goals of those designs. Many applications (e.g. tailoring focal length, beam steering and shaping, polarization-based routing and others), require flexible control over a broader span of phase values, covering the range $[-\pi, \pi]$, including abilities to design spatial discontinuities in parameters. This functionality, however, strongly depends on the illumination frequency. Broadband operation usually requires cascading several systems. Here, this general approach of functional multiplexing in order to focus two different polarizations into a pair of different spatially separated focal points will be undertaken.

A functionality of focal positions discrimination for mutually orthogonal polarizations, is demonstrated (Fig. 1). Polarization-sensitive two-dimensional pattern on a single printed-circuit-board (PCB) is implemented and optimized towards spatial separation of focal spots. The device operates in a broad spectral range, demonstrating over a wavelength separation distances and enables simultaneous acquisition of cross-polarized RF images.

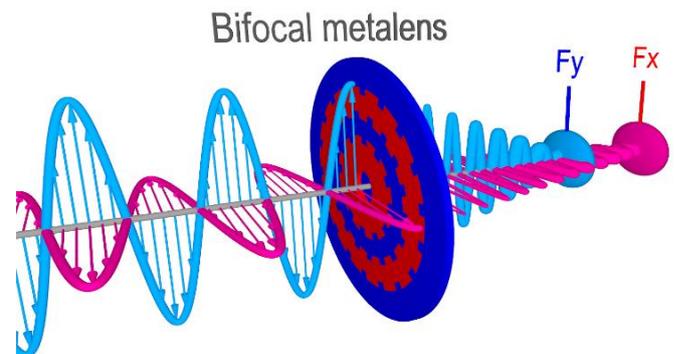

Fig. 1. Bifocal Fresnel zone plate lens - two orthogonal polarizations are focused at different points along the propagation axis owing to interactions with polarization-sensitive overlaping metasurface-based Fresnel zones.

The letter is organized as follows: the design concept is followed by numerical and experimental results. Comparison with standard Fresnel lens formulas comes before the concluding remarks and the outlook.

## II. BIFOCAL LENS DESIGN

The bifocal lens design utilizes the basic Fresnel zone concept. A set of alternating opaque-transparent concentric rings, either blocking or transmitting Fresnel zones, are defined with the following set of radii:

$$r_n = \sqrt{n\lambda F + \frac{n^2\lambda^2}{4}}, \quad (1)$$

where $n$ is the number of the zone, $\lambda$ is the wavelength that the lens is designed for, $F$ is the focal distance of the lens and $r_n$ is the outer radius of the $n$-th ring, also corresponding to the inner radius of the $n+1$-th ring. Relation (1) explicitly shows that the Fresnel radii depend on the focal distance and, hence, in order to achieve multiple focal points several types of zones should be overlapping. It can be realized by stacking multiple zone plate lenses one on top of another or by using the polarization degree of freedom of electromagnetic waves – this approach will be followed hereafter.

A basic scheme of the lens is shown on Fig. 1 – a flat structure provides two distinguishably different focal points along the optical axis of the system. Polarization-sensitive response of a structured surface is achieved by designing elongated patches, resonant with one pre-defined polarization. The orthogonal component of the field weakly interacts with short axis of a patch and, as a result, polarization-dependent opaqueness or transparency is achieved. At the region, where Fresnel zones for both polarizations overlap (recall the different focal distances influence), orthogonal patches will create a cross-shaped planar antennas. Here, a careful optimization design is needed in order to prevent cross-polarization coupling.

The lens was designed to operate at 10 GHz central frequency and to provide two focal spots for different polarizations located at $F_y$=90mm (Y-lens) and $F_x$=150mm (X-lens) from the patterned surface - the focal points are separated by two wavelengths. Fig. 2(a, b) schematically show the positions of the focal spots depending on the polarization of the incident field. Such separation should result in two distinguishable and measurable spots in the field distribution. Y-lens is consisted of 6 zones (filled even zones) and X-lens was made of 4 zones (filled even zones) (Fig. 2(d)). The relevant parameters are summarized in Table 1.

| $r_n$ | X Polarization | Y Polarization |
| --- | --- | --- |
| $r_1$ | 68.74mm | 54.08mm |
| $r_2$ | 99.5mm | 79.37mm |
| $r_3$ | 124.6mm | 100.62mm |
| $r_4$ | 146.97mm | 120mm |
| $r_5$ | -- | 138.29mm |
| $r_6$ | -- | 155.88mm |

Table 1. Zone plates parameters.

An optimization of the metasurface properties was made in order to reduce polarization crosstalk and minimize sizes of focal spots (depth of focus- in Z direction). The metal patches dimensions were selected to be $\lambda/5$ by $\lambda/20$, which yielded the values of 6 and 1.5mm correspondingly for 10 GHz frequency (Fig. 2(c)). The gap between the two adjacent rods was chosen to be 1.5mm. The total size of each metasurface cell was taken to be 7.5mm by 7.5mm. Fig. 2(c) shows the design of vertical, horizontal, and overlapping regions. Electromagnetic properties of substrates, used in the experimental demonstration (copper-plated PCB, copper thickness – 100 micrometers, substrate – 1.5mm of FR-4 glass epoxy), were introduced for obtaining the optimized design.

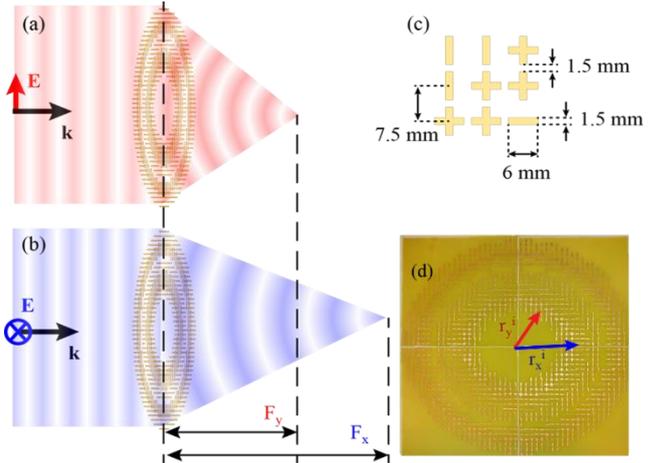

Fig. 2. Bifocal Fresnel lens – the concept and design. (a), (b) Schematics of bifocal operation – orthogonal linear polarizations are focused at different distances from the flat patterned surface. (c) Single cell's characteristic dimensions: horizontal, vertical, and cross-shaped patches. (d) A photograph of the fabricated lens - four PCB plates form a 328mm×328mm flat lens.

## III. RESULTS

Performance evaluation and optimization of the lens design were performed numerically with CST Microwave studio ("Time Domain Solver", ~ 2.4 mesh cells per $mm^3$). The lens was fabricated with a help of photolithography, followed by etching of copper-plated PCBs. A wideband horn antenna, connected to the transmitting port of a vector network analyzer (VNA, Agilent E8362B), was used as a plane wave excitation source. The lens is located at the far-field zone of the antenna (2.5 m apart). The fields were measured with a 3D near-field scanner. An electrical field probe was mounted on the scanner and connected to the receiving port of the VNA. The shielded aperture of the probe has a 2 mm diameter and could be approximated as an electrically small perturbation that measures electrical fields along a given direction, without performing a significant field averaging. In the case of study here, the probe was oriented normally to the incident $k$-vector and parallel to the dominant electric component of the field, radiated from the horn. The near-field mapping is performed over a rectangular area (250x350 mm) with 3 mm steps ($\lambda/10$) in both (Y and Z) directions starting at 2 mm offset from the back interface of the lens, in order to avoid a contact between the probe and the sample.

Fig. 3(a, b) shows numerically obtained electric field intensity distributions for two different polarizations of the incident plane wave. Two well distinguishable focal spots can be observed and follow the initial design requirements. It can be observed that X-polarized wave shows larger depth of focus (DOF) - the focal spot is extended along the optical axis. This effect is attributed to the different focal length and different number of Fresnel zones, used for X- and Y- polarizations (150mm with four zones and 90mm with six zones, correspondingly), as will be discussed later. Furthermore,

Fresnel zone plates create a back reflected focal spot, which is situated at the same distance as the forward focus, but in the negative direction along the optical axis. Numerical modeling (not shown here) clearly indicates its existence.

The numerical data can be directly compared with the experimentally obtained near field scans in YZ plane (Fig. 3(c, d)). While the same focal positions were obtained in both numerical and experimental cases, the latter one demonstrates non-uniform (wavy) distribution of the field in the focal spot. The clearly distinguishable periodicity corresponds to the wavelength of the excitation and is quite typical for experiments, performed in an anechoic chamber. Additional reflections from excitation, collection horns and metallic parts of the 3D scanner create those interference patterns. Residual field distribution in space originates from the finite number of Fresnel zones in the fabricated lens that do not block the entire incident wave.

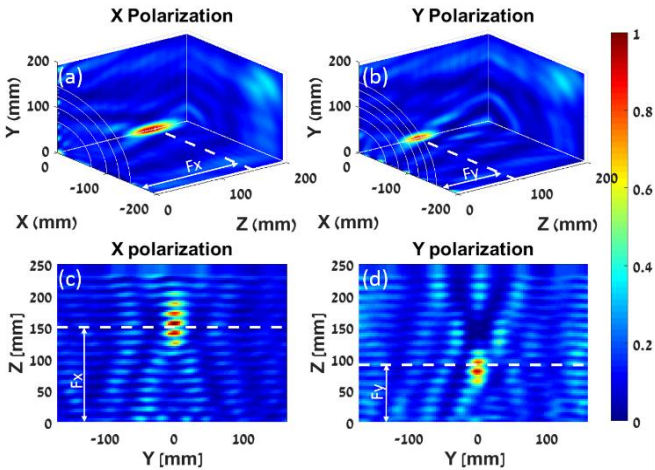

Fig. 3. Transmitted electric field intensity distributions after the bifocal Fresnel lens, illuminated with 10 GHz plane wave. White dashed lines indicate positions of the highest intensity (focal position along the optical axis). Numerical data for: (a) X-polarized incident wave, (b) Y-polarized incident wave. Electric field intensity distribution after the lens in YZ plane – experimental data: (c) X-polarized incident wave, (d) Y-polarized incident wave

Eq. 1 clearly demonstrates a strong chromatic dependence of the focal positions. Nevertheless, spatial separation of focuses for orthogonal polarizations still holds for different frequencies of the incident wave. This bifocal performance in the range of 8-12 GHz will be tested next. The results of the frequency sweep measurements are presented in Fig. 4, which demonstrates the intensity scans along the optical axis for the chosen frequency range (each vertical line states for one scan per frequency). As it can be observed in both the numerical and the experimental investigations, the focal positions Fx and Fy do move as the function of frequency, nevertheless spatial separation between them is preserved. Experimental data suffers from the same wavy behavior, observed in the 3D field scans (Fig. 3(c,d)). The slopes of the white dashed lines (main focal positions) in both of the cases are nearly identical, which results in a constant distance between X- and Y- focal spots (~60mm). Higher-order focuses, located farther from main focal spot, could be distinguished.

The shift of the focal position with the frequency change is given by the next approximation [23]:

$$F(\omega_{incident}) \cong F_{designed}\left(2 - \frac{\omega_{designed}}{\omega_{incident}}\right), \quad (2)$$

where $F_{designed}$ and $\omega_{designed}$ are the focal distance and the frequency respectively that the lens designed for. $\omega_{incident}$ is the actual frequency of the incident plane wave, and $F(\omega_{incident})$ is the calculated/measured focal distance of the lens for incident wave with frequency $\omega_{incident}$.

The depth of focus could be approximated by [23]:

$$DOF \cong \pm \frac{F(\omega_{incident})}{2N}, \quad (3)$$

where DOF is the depth of focus ($\pm$ stays for distances spread from the central focal point) N is the number of zones related to the wave polarization. In the bifocal design both F and N are different for orthogonal polarization.

Both of (2) and (3) are valid approximations for the case, when $N\lambda F \gg \left(\frac{N\lambda}{2}\right)^2$ (recall the expressions in (1)). This condition is satisfied relatively well in the current design (factor of ~3 in the ratio). The parameters if the bifocal lens are summarized in Table 2. Values, calculated with (2) and (3) along with those, extracted from actual measurements (those numbers appear in round brackets) can be directly compared. It can be seen that the results agree relatively well and the differences are in the range of ~10% at frequencies, where initial design was performed.

| $\omega_{incident}$ | X Polarization – (N=4) | | Y Polarization – (N=6) | |
|---|---|---|---|---|
| | $F(\omega_{incident})$ | DOF | $F(\omega_{incident})$ | DOF |
| 8GHz | 112.5mm (~110mm) | ±14.06mm (~±16mm) | 67.5mm (~60mm) | ±5.63mm (~±7mm) |
| 10GHz | 150mm (~155mm) | ±18.75mm (~±21mm) | 90mm (~90mm) | ±7.5mm (~±9mm) |
| 12GHz | 175mm (~200mm) | ±21.88mm (~±24mm) | 105mm (~115mm) | ±8.75mm (~±10mm) |

Table 2. Summary of the bifocal lens parameters – theoretical values appear at the upper rows of each cell, while experimental data is enclosed in round brackets.

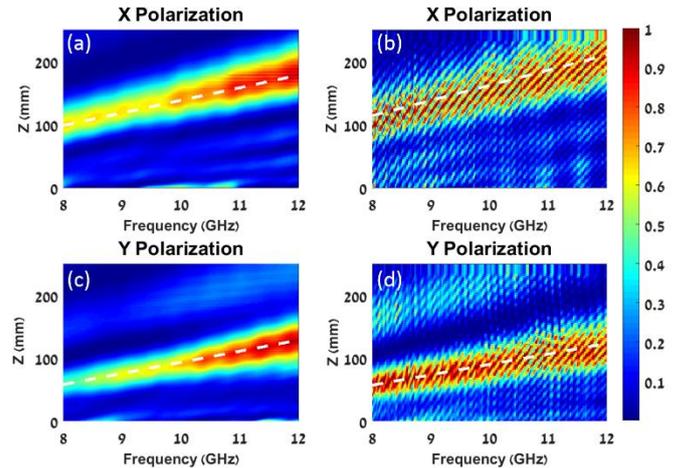

Fig. 4. Electric field intensity distribution along the optical axis of the lens (vertical lines) for frequencies range between 8GHz and 12 GHz (horizontal axis). The color maps are normalized to the same value. The white dashed lines follows the maximum of the field intensity (the focal spot). Numerical data for: (a) X polarization, (c) Y polarization. Experimental data for: (b) X polarization, (d) Y polarization.

IV. CONCLUSIONS

The approach towards creation of polarization-sensitive bifocal lens has been demonstrated both numerically and experimentally. The Fresnel zone plate lens has been used as a basis for the studies. It was demonstrated, that the zone plate lens consisting of sub-wavelength metal patches yields polarization sensitivity of the focal spot position. The versatility of the approach was supported by the studies of the lens performances in the wide frequency range (8-12GHz). The proposed lens design is susceptible to chromatic aberration, though it provides the stability in relation between focal positions for different polarizations, which might be useful for polarization state mapping and RF imaging in orthogonal polarizations at the same time. Similar design rules could be used for demonstration similar performances at optical spectral range. Furthermore, RF designs could serve as emulation tools for studies of complex photonic phenomena, where both fabrication and characterization are resource-consuming, e.g. [24], [25], [26].

ACKNOWLEDGMENTS

This work was partly supported by TAU Rector Grant, PAZY foundation, Kamin project.